\documentclass[11pt, oneside]{article}   	
\usepackage{geometry}                		
\usepackage{graphicx}				
\usepackage{amssymb}
\usepackage{algorithm}
\usepackage{algpseudocode}
\usepackage{hyperref}

\algnewcommand\algorithmicforeach{\textbf{for each}}
\algdef{S}[FOR]{ForEach}[1]{\algorithmicforeach\ #1\ \algorithmicdo}

\usepackage{multirow}
\usepackage{longtable}
\usepackage{authblk}

\title{Evaluating the Risk of Changes in a Microservices Architecture}
\author[1]{Matteo Collina}
\author[2]{Luca Maraschi}
\author[3]{Tommaso Pirini}
\affil[1,2]{Platformatic Inc.}
\affil[1,2]{\textit{\{matteo,luca\}@platformatic.com}}
\affil[3]{\textit{me@tommasopirini.com}}

\begin{document}

\maketitle

\begin{abstract}
In a microservices-based system, reliability and availability are key components to guarantee the best-in-class experience for the consumers. One of the key advantages of microservices architecture is the ability to independently deploy services, providing maximum change flexibility. However, this introduces an extra complexity in managing the risk associated with every change: any mutation of a service might cause the whole system to fail.
In this research, we would propose an algorithm to enable development teams to determine the risk associated with each change to any of the microservices in the system.
\end{abstract}

\section{Introduction}
Monolithic Architecture has long been a prominent and widely used software development approach. This architectural paradigm is characterized by bundling all a software application's components, modules, and functionalities into a single, colossal codebase or repository. While the monolith architecture has its merits, such as simplicity in deployment and development, it also presents many challenges, including the intricate web of connections between team members that grows exponentially as a project expands.
\par
Imagine a software project with `n' developers, each working on different parts of the monolithic codebase. In such a scenario, the number of possible connections, or communication paths, between these team members is determined by the formula \(n*(n-1)/2\) \cite{hackman1983normative}. This formula illustrates that the number of potential communication links between developers escalates exponentially as the team size increases. While collaboration is essential for a project's success, this intricate network can lead to various problems, ranging from increased communication overhead to decision-making bottlenecks and even a loss of focus on individual responsibilities.
In today's dynamic and rapidly evolving technological landscape, microservices architecture has emerged as a transformative software development and deployment approach \cite{zimmermann2017microservices}. 
\par
Microservices architecture 
\cite{alshuqayran2016systematic,dmitry2014micro,li2021understanding,de2019monolithic} is a technical solution that effectively addresses the operational complexity of growing an engineering organization. It accomplishes this by solving what can often be perceived as a ``people problem" – the challenges that arise when multiple teams collaborate on a single application. Here's a detailed explanation of how microservices mitigate operational complexities and enable parallel work with minimized interactions:
\begin{itemize}
\item \textbf{Decomposition of Monoliths}: In a traditional monolithic application, all components are tightly coupled within a single codebase. As the application grows, this complexity compounds. Microservices, on the other hand, advocate breaking down the monolith into smaller, independent services, each responsible for a specific set of functionalities. This decomposition isolates complexity within each microservice, making it more manageable and understandable.
\item \textbf{Autonomous Development and Deployment}: Microservices allow teams to own and manage individual services independently. Each team is responsible for developing, testing, and deploying their microservice. This autonomy reduces the need for cross-team coordination, as teams can make decisions within their service's scope without affecting other teams. It enables faster development cycles and promotes a sense of ownership among teams.
\item \textbf{Parallel Development}: Microservices enable concurrent development by multiple teams on different services. This parallelization is crucial for scaling an engineering organization because it eliminates bottlenecks caused by teams waiting to access a shared monolithic codebase. Each team can work on its microservice without being blocked by other teams, accelerating feature delivery.
\item \textbf{Minimized Interactions}: In a monolithic environment, the complexity of communication and coordination increases as the number of developers and teams increases. Microservices mitigate this problem by limiting interactions to well-defined Application Programming Interface (API) contracts between services. Teams can work in isolation as long as they adhere to these contracts. This minimizes the need for extensive cross-team discussions, reducing operational overhead.
\item \textbf{Fault Isolation}: In a monolithic system, a single bug or failure can have cascading effects across the entire application. Microservices, by design, contain faults within the scope of a single service. If one microservice encounters an issue, it is less likely (but how much?) to affect others. This fault isolation reduces the blast radius of problems, making it easier to manage and troubleshoot issues.
\item \textbf{Scalability}: Microservices allow organizations to scale services independently based on demand. This elasticity is vital as it ensures that resources are allocated efficiently, reducing the operational complexities associated with resource management in a monolithic system.
\item \textbf{Technological Diversity}: Different microservices can use different technologies and languages that best suit their requirements. This flexibility accommodates engineering teams' diverse skill sets and preferences, eliminating the need for one-size-fits-all technical decisions.
\end{itemize}

In summary, microservices address the operational complexity of scaling an engineering organization by breaking down monolithic applications into manageable, autonomous services. This technical solution enables multiple teams to work in parallel with minimal interactions, reducing coordination overhead and allowing each team to focus on its area of expertise. By isolating complexity, promoting autonomy, and facilitating fault isolation, microservices offer an effective approach to handling the "people problem" in growing engineering organizations.
The promise of increased agility, scalability, and continuous delivery has led organizations worldwide to embrace microservices as a fundamental building block of their IT ecosystems \cite{hackman1983normative}. However, beneath the surface of this enticing paradigm shift lie common challenges that agile and distributed teams face when developing microservices.
\par
The allure of microservices often conceals the intricacies of managing these granular, interdependent components. Agile teams, renowned for their flexibility and adaptability, can grapple with the complexities of coordinating and deploying numerous microservices effectively. Distributed teams operating across geographical boundaries face additional hurdles in communication, synchronization, and maintaining a cohesive development process, while having to coordinate changes across shared APIs without causing disruptions.
\par
Change is a constant in software development and a fundamental part of daily operations in microservices. Each modification, update, or addition to a microservice can have far-reaching implications, affecting the entire system's stability, performance, and reliability. As organizations strive to meet the demands of an ever-evolving market, it is imperative to proactively address these challenges and establish a systematic approach to assess and manage the risks associated with changes in a microservices environment. The state of the art solutions to address this problem range from cross-team API design sessions to black-box integration testing, to extensive manual quality assurance sessions. Those are a significant investment to enact, and they significantly slow down development.
\par
In this paper, we will delve into the intricacies of microservices and the associated risks of change. We will explore the limitations of traditional risk assessment methods in this context and underscore the need for a dedicated approach. Our system's architecture and key components will be detailed, illustrating how it captures, analyzes, and quantifies change-related risks. Additionally, real-world case studies and practical examples will showcase the tangible benefits and results organizations can achieve by adopting this innovative approach.
\par
In this paper we are going to explore the solution for a common scenario, in which multiple development teams are developing and committing changes to individual applications’ repositories, with both loose and strongly coupled dependencies (ex. a service calling an endpoint on another service on the network).

\subsection{Glossary}

This is a section that explains technical words used here, as we have interpreted them to best represent the objects of this work:
\begin{itemize}
\item \textbf{\textit{Branch}}: part of a path, starting from the entry-point and ending in a microservice that no longer generates API requests.
\item \textbf{\textit{Breaking change}}: alteration of an operation that causes an API request to fail, requiring a change in the API client.
\item \textbf{\textit{Client}}: user calls operations from microservices.
\item \textbf{\textit{Entry-point}}: first point of contact to the microservices system from which the user makes requests.
\item \textbf{\textit{Microservice}}: software entity interacting with other software entities via message passing communications using standard data formats and protocols (e.g., XML, SOAP and HTTP) and well-defined interfaces \cite{dragoni2018microservices}.
\item \textbf{\textit{Operation}}: execution of a request via API to a microservice.
\item \textbf{\textit{Path}}: route that calls API operations to microservices to perform a task for the end user.
\item \textbf{\textit{Risk}}: possibility of a client failure when calling to an API operation.
\end{itemize}

\section{Related Work}

In real-world software systems, there are many issues with knowing which activities will be broken by breaking changes in APIs and how they will affect the entire system. Some papers analyze behaviors and historical data about changes in software APIs and libraries, but only a few of them are trying to address the need to know as early as possible the effects of breaking changes in the system.
\par
In \cite{xavier2017historical} the authors present a study to measure the amount of API breaking changes on real-world Java libraries and its impact on clients at a large-scale level, and in \cite{brito2018apidiff} they introduce APIDIFF, a tool to identify API breaking and non-breaking changes between two versions of a Java library using a combination of similarity heuristics and static analysis of Java source code. Our researchprovides instead  a risk index that can be used to foresee the impact of any breaking change before committing the change to the system. 
In these studies, as \cite{dig2006apis}, the authors distinguish API changes in breaking changes and non-breaking changes, but to evaluate the risk of failures here we don’t consider the last ones, because they will not condition the chance to break user activities.
\par
Organizations usually are concerned about managing risk, i.e., the likelihood that a code change will cause a negative impact on their products and processes. In \cite{shihab2012industrial} the authors show different criteria for determining risky changes, using factors extracted from the changes and the history of the files modified by the changes, such as the number of lines and chunks added by the changes, the bugginess of the files being changed, the number of bug reports linked to the change and the experience of the developer making the change. However, these techniques look for the probability of risk based on the developer and team level, and do not provide an analysis based on a modern microservices system, as we propose in our work.
\par
In \cite{ochoa2022breakbot}, the authors introduce BreakBot, a bot that analyzes the pull requests of Java libraries on GitHub to identify the breaking changes they introduce and their impact on client projects. Through static analysis of libraries and clients, it extracts and summarizes objective data that enrich the code review process by providing maintainers with the appropriate information to decide whether and how changes should be accepted, directly in the pull requests. While this work focuses on source- and binary-incompatible changes and leaves behavioral changes aside, our work instead considers a more general approach, not focused on a programming language, such as Java, but providing an easy-to-read risk index for any type of breaking changes in microservices systems.
\par
In \cite{rohani2014calculating}, the authors propose a method to enable the IT management team of KLM to predict their Business Application availability based on the configuration and the components used in their infrastructure. In fact they state calculating the availability of a Business Application within a very complex organization is not easily achievable. We started from this approach to generalize the calculation of the risk of failure on any real-world microservices system.
\par
In our experiments, we used a vectorial data representation of microservices’ activities that can show the situation of the system in a time snapshot. Nowadays, there are various technologies to extract and manage this kind of data, we used OpenTelemetry \cite{opent}.

\section{Algorithm}

In this section we describe the algorithm that computes the risk value for the end user to make a breaking change set in operations made by microservices. 
\par
This paper aims to define an algorithm that calculates a failure risk value of a software system, and not to define the dependency graph between the various components of the system itself. Therefore, whether there is an operation in the system that is never requested by any user activity, in the analyzed time snapshot, this will not be considered as it is not used to calculate the system risk. In fact, the risk of an operation that is never requested is zero by definition, i.e. it represents a path that is never crossed. We remember that we calculate the risk on real data of a general software system, not on a static representation of the dependencies between the elements of the system.

\subsection{Definitions and assumptions}

First of all, we describe the components of our system and the assumptions we make for the reliability of the proposal. Therefore we consider a microservice architecture because of its independent deployability and lightweightness and its application in the cloud \cite{pahl2016microservices, dragoni2018microservices, waseem2021design}.
In particular, our system is composed of a specified number of microservices. Each of them provides some operations, called by others a number of times.
\par
End users start to interact with the system by an entry-point, calling an operation provided by a microservice. Each microservice to serve a request can call another operation provided by another microservice, and so on, composing a path of operations. Each path is independent with respect to the related microservices. Whenever an operation invokes multiple operations in other microservices, the path is branched out depending on the complexity expressed by the system activities. The amount of end users' activities are also considered comparable, which means that here we don't consider users to be much more active than others, or a few users more “important” than others. This is actually what usually happens in this kind of software system. All paths are monitored and this information is collected and stored in a database. Lastly, our observation point is the entry-point of the system, where all the call paths are started by users.
\par
We aim to calculate the end user risk value between 0 and 1, where 0 represents no risk, which means there is nothing broken in the system, and 1 represents that all paths in the system are at risk, which with a high likelihood might cause the system to fail. We choose this range of values because the risk is represented as the difference between the situation of the system as result of a breaking change and the ideal situation of the system, when every operation works. We inspire this choice at the Jaccard index or Jaccard similarity coefficient, that is a statistic used for gauging the similarity and diversity of sample sets \cite{jaccard1912distribution}.
\par
In this paper we assume as well that each microservice has the same importance, so we don't give more importance to a microservice with respect to others, we don't manage for now this kind of information.
Furthermore, in our representation of the call paths it is also possible to foresee circular calls, which however will always be resolved not in a loop, due to the real-world nature of our systems: sooner or later a microservice must return an answer. So, we want to consider a linear system without loops, and we model our representation mode according to it.
\par
\begin{figure}
  \includegraphics[width=\linewidth]{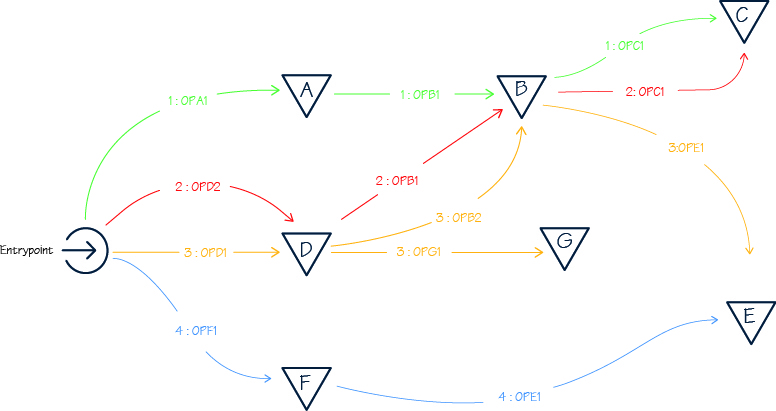}
  \caption{An example graph figured the testing system.}
  \label{fig:graph}
\end{figure}

In Figure \ref{fig:graph} we have a visual representation of a simple system example. In particular, we have an entry-point and seven microservices, called from A to F. The operation calls among them are defined as OP followed by the name of the microservice supplier and an ordered number: e.g. $OPA1, OPD2$.
For example, the path number 3 that starts from the entry-point that calls microservice D using $OPD1$, moreover D calls microservice B by $OPB2$ and microservice G by $OPG1$, and finally B calls microservice E using $OPE1$.

\par
Each connection stores the number of operations called between the microservices, in a specific time snapshot of the system. So, for example, we have the following entry in the Microservice Path database (MSP):

\[ OPD1;OPB2 = 34 \]

that means, in this path, the operation OPB2 requested by microservice D is called 34 times.
\par
The data structure that stores this information is a key-value database where for each path numbered, we save the number of calls of each step of the path. For example, in path number 3 we store how many times $OPD1$ is called by the entry-point, $OPB2$ is called by D, $OPG1$ is called by D, $OPE1$ is called by B. In the end, we store all the calls of the systems.
For example, a path like this:

\[ p3 = [ OPD1: 45, OPD1;OPB2: 34, OPD1;OPG1: 24, OPD1;OPB2;OPE1: 20 ] \]

\subsection{Implementation}
With this data, we determine the algorithm \ref{alg:risk} that computes a value representing the risk that a list of operations’ breaking changes reach the entry-point of the system, failing users' activities.

\begin{algorithm}
\caption{Algorithm to obtain the risk value from a list of operations’ breaking changes}
\label{alg:risk}
\begin{algorithmic}
\Require The Microservice Paths database $MSP$
\Require A set of breaking change operations $S=[OPX_1, OPX_2…OPX_N]$
\State $r \gets 0$
\ForEach{$p$ in $MSP$}
   \ForEach{$o$ in $S$}
      \ForEach{$t$ in $p$}
         \If{$o$ is contained in $t$}
            \State $r \gets$ r + $risk$($t, o$)
         \EndIf
      \EndFor
   \EndFor
\EndFor
\end{algorithmic}
\end{algorithm}

In Algorithm \ref{alg:risk}, we require the Microservice Path database that stores all the call paths with respective amounts in a time snapshot of the system, and the set of breaking change operations $S$, that can not be empty, but can contain a single operation or many of them, belonging to a single microservice, to a software monolith or to various microservices.
\par
The starting value for the risk is zero, because it's the neutral value of the addition, and the final risk value $r$ is computed as the sum of the risk value of the paths calling operations in set $S$. Therefore, Algorithm \ref{alg:risk} browses all paths in MSP, and for each path $p$ it checks if almost one of the operations in the breaking change set $S$ is contained in path $p$. To do that the algorithm analyzes all the branches of the path. A branch $t$ of the path is the operation call made by a microservice to another one, that stores a number representing the amount of calls made at that branch, until the snapshot. When the branch t contains the operation o we increment r with the result of the risk function, computed as:

\[ risk(t, o) = \sum_{i=1}^{n} \frac{req(ms_i,OPX_i)}{req(ms_i)} * c_{ms_i} \]

where the result of the risk formula called on the branch $t$ and the operation $o$ is the sum of all the risk values of the operations of the branch $t$, from the first one, usually made by the entry point, to the last one $n$. A branch consists of an operation called in sequence. The risk value of the $i$-th operation is computed by the fraction between the number of the $OPX_i$ requests made by the microservice $ms_i$ in the branch $t$, and the number of all the requests made by the microservice $ms_i$ in every path of the system. This fraction besides is multiplied by a coefficient $c_{ms_i}$ related to the microservice $ms_i$, and computed as:

\[ c_{ms_i} = \frac{req(ms_i)}{req(MSP)} \]

where $req(MSP)$ is the number of total requests in the system.
\par
Both the fraction values and the coefficient values are between 0 and 1. This assumption is made because the numerator is always less than or equal to the denominator, being a subset of the total calls to the microservice, for the first one, and of the entire system, for the second one.
The coefficient is used to weigh the operation call with respect to the entire system, therefore the sum of all coefficients should be 1. Whether an operation of a microservice is called a lot of times, the risk of breaking it must be high, otherwise whether an operation of a microservice is called a few times, the risk must be low.
\par
With the incorporation of this coefficient, the risk computation stands to be streamlined, and the dependency on $req(ms_i)$ can be eliminated. Nonetheless, this paper primarily focuses on elucidating our solution's rationale, deferring the fine-tuning of parameters and formulas to future research endeavors.

\section{Experimental results}

In this section, we show the results of the execution of our algorithm on the example scheme of Figure \ref{fig:graph}, suitably supported by random numerical values for each path operation.
\par
We consider four different scenarios to represent a full view of reality. The first scenario is the simplest and consists in the situation where there is a single breaking change operation in a microservice. The second scenario, more realistic, also considers a single microservice in which, however, there are multiple breaking change operations. The third scenario contemplates the case of an application with a breaking change set on the critical path. The last scenario assumes that in a microservices system, more microservices can be changed in multiple operations over time. They don't change all microservices at the same time but in a time sequence. These scenarios are not considered only at runtime but in a global way. In our analysis we deal with time snapshots, so we don’t have the problem of time sequence.
\par
Moreover, we considered three use cases that diversify random values. The first two use cases are quite similar, to see how the algorithm behaves in comparable situations. The third use case, on the other hand, differs by increasing the values of one path, the first, by an order of magnitude, with respect to the others.
The values used in these use cases are in Table \ref{tab:example}.

\begin{longtable}[c]{| c | c | l | c |}
\caption{Example value for three use cases.\label{tab:example}}\\
 \hline
 \textbf{MSP} & \textbf{Path number} & \textbf{Path branch} & \textbf{Value}\\
 \hline
 \endhead

 \hline
 \endfoot

\multirow{12}{13em}{mce0} & \multirow{3}{4em}{1} & $OPA1$ & 32 \\
& & $OPA1:OPB1$ & 20 \\
& & $OPA1:OPB1:OPC1$ & 18 \\
\cline{2-4}
& \multirow{3}{4em}{2} & $OPD2$ & 65 \\
& & $OPD2:OPB1$ & 44 \\
& & $OPD2:OPB1:OPC1$ & 41 \\
\cline{2-4}
& \multirow{4}{4em}{3} & $OPD1$ & 23 \\
& & $OPD1:OPB2$ & 20 \\
& & $OPD1:OPG1$ & 12 \\
& & $OPD1:OPB2:OPE1$ & 15 \\
\cline{2-4}
& \multirow{2}{4em}{4} & $OPF1$ & 52 \\
& & $OPF1:OPE1$ & 43 \\
 \hline
 
 \multirow{12}{13em}{mce1} & \multirow{3}{4em}{1} & $OPA1$ & 60 \\
& & $OPA1:OPB1$ & 42 \\
& & $OPA1:OPB1:OPC1$ & 42 \\
\cline{2-4}
& \multirow{3}{4em}{2} & $OPD2$ & 23 \\
& & $OPD2:OPB1$ &18 \\
& & $OPD2:OPB1:OPC1$ & 14 \\
\cline{2-4}
& \multirow{4}{4em}{3} & $OPD1$ & 59 \\
& & $OPD1:OPB2$ & 48 \\
& & $OPD1:OPG1$ & 44 \\
& & $OPD1:OPB2:OPE1$ & 33 \\
\cline{2-4}
& \multirow{2}{4em}{4} & $OPF1$ & 21 \\
& & $OPF1:OPE1$ & 20 \\
 \hline

 \multirow{12}{13em}{mce2} & \multirow{3}{4em}{1} & $OPA1$ & 400 \\
& & $OPA1:OPB1$ & 340 \\
& & $OPA1:OPB1:OPC1$ & 290 \\
\cline{2-4}
& \multirow{3}{4em}{2} & $OPD2$ & 38 \\
& & $OPD2:OPB1$ & 36 \\
& & $OPD2:OPB1:OPC1$ & 27 \\
\cline{2-4}
& \multirow{4}{4em}{3} & $OPD1$ & 40 \\
& & $OPD1:OPB2$ & 36 \\
& & $OPD1:OPG1$ & 28 \\
& & $OPD1:OPB2:OPE1$ & 10 \\
\cline{2-4}
& \multirow{2}{4em}{4} & $OPF1$ & 34 \\
& & $OPF1:OPE1$ & 29 \\
   
 \end{longtable}

\begin{figure}
  \includegraphics[width=\linewidth]{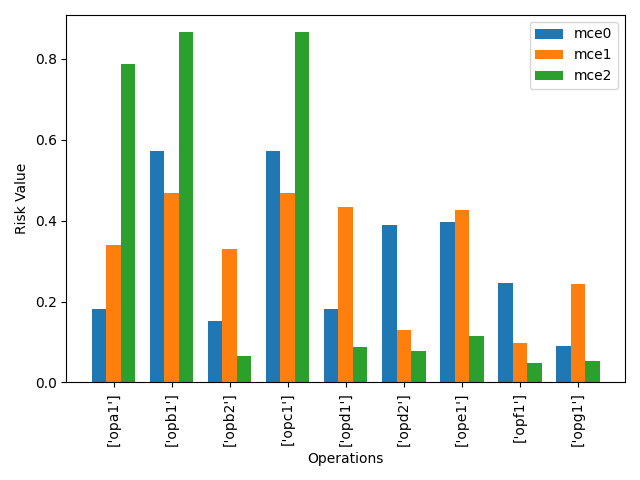}
  \caption{Experimental results with single breaking change in operations.}
  \label{fig:res1}
\end{figure}

The algorithm is run in several cases. In Figure \ref{fig:res1} we have the comparison between the results of the risk values, on the y-axis of the graph, that are obtained when only one operation of the system at a time is broken. We have considered all single operations, which are on the x-axis of the graph, in the three case scenario, on the legend, with different colors.

We can see that the results of the first two MSPs ($mce0, mce1$), with comparable data, have more or less the same values, more specifically they differ by a little more than 20\% at most. While for the MSP $mce2$ (the green one), which has path 1 with values of an order of magnitude greater, it differs by significantly increasing the risk value when the operations affected by the breaking changes are in path 1, and considerably lower in the other cases. This shows that if there are paths with more user activity, a breaking change in these paths affects the risk value more.

\begin{figure}
  \includegraphics[width=\linewidth]{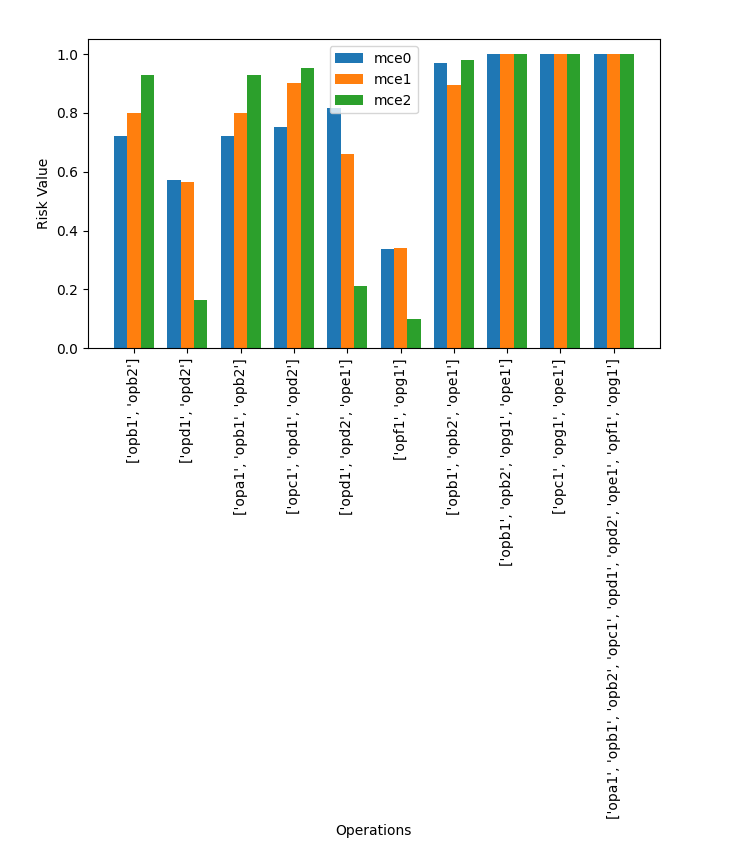}
  \caption{Experimental results with many breaking changes in operations.}
  \label{fig:res2}
\end{figure}

In Figure \ref{fig:res2} we show instead the experimental results concerning many operation breaking changes, relating to the same microservice or on different microservices. In this case we notice that whether entire microservices fail, with more than one operation used in different paths, the risk is significantly higher, and is often close to 1. This means that the risk of system failure is higher and will therefore affect user activity.
In some cases, there is also an evident difference among the results of the various MSPs, in particular with $mce2$, where it is always noted that if the operations affected by breaking changes are those of path 1, the risk is greater, because they have many requests than other paths, and the risk is lower, when it includes unused operations in path 1.
\par
The last three evaluations, on the right of Figure \ref{fig:res2} , consider cases where all paths of the MSP have at least one breaking change. The result is that every user activity will be affected by failures, therefore the risk of the system is 1, i.e. you can have almost certainty that the user experience will be affected, and the other services would need to be updated. The last case, specifically, highlights what happens when all the operations of microservices have breaking changes, and also in this case the risk value is 1, because it means that all operations have been affected by the attempted changes.
\par
These experimental results, carried out on the real data shown in Table \ref{tab:example}, demonstrate that the execution of the proposed algorithm provides a useful index to understand how much impact a breaking changes set has in the analyzed software system.

\section{A possible real-life implementation}

In Platformatic’s implementation of this algorithm, we have harnessed the power of OpenTracing to collect invaluable real traffic data. OpenTracing, an essential component of the OpenTelemetry \cite{opent} project, offers a robust framework for observability in distributed systems. It enables us to gain deep insights into our application's performance, troubleshoot issues, and optimize efficiency by collecting and analyzing data from various microservices and components. To fully grasp our implementation, it's crucial to understand the key concepts of OpenTracing, which include traces, spans, and distributed trace, as well as how we utilized the ``span-kind" feature to detect and track calls between microservices.

\subsection{Understanding Traces, Spans, and Distributed Traces in OpenTelemetry}

\begin{enumerate}
\item \textbf{Traces}: Traces are the fundamental building blocks of observability in OpenTelemetry. They represent the entire journey of a request or transaction as it traverses through various components of a distributed system. Think of a trace as a chronological sequence of events that capture the lifecycle of a request, allowing us to see where time is spent and how different parts of our application interact. In a more specific definition, a trace is a direct acyclic graph of spans.

\item \textbf{Spans}: Spans are individual units of work within a trace. They encapsulate a specific operation or task, such as processing an HTTP request, querying a database, or invoking a microservice. Spans are used to measure the duration of these operations, record context, and provide essential contextual information, like error status or tags, to help us understand what happened during each operation. A ``root span" represents in our case the first call from the entrypoint.

\item \textbf{Distributed Trace}: A distributed trace is a collection of related spans across multiple services that collaborate to fulfill a single user request. Distributed tracing is pivotal for identifying bottlenecks, diagnosing latency issues, and understanding the flow of requests as they traverse through various microservices and components in a distributed architecture.
\end{enumerate}

\subsection{Utilizing Span-Kind for Microservices Communication}

One of the key challenges in distributed systems is tracking and monitoring communication between microservices. This is where the ``span-kind" feature in OpenTelemetry comes into play. Span-kind allows us to categorize spans based on their role in a transaction. Two common span kinds are:

\begin{itemize}
\item[-] \textbf{Client Span}: This represents the initiating span for an outbound request from one microservice to another. It captures the time taken for the request to travel to the target service and receive a response.

\item[-] \textbf{Server Span}: On the receiving end, the server span captures the processing time of the inbound request. It provides insights into how the microservice handles the request and its response time.
\end{itemize}

By utilizing span-kind, we can precisely detect and track calls between microservices. This granular visibility empowers us to identify performance bottlenecks, troubleshoot issues, and optimize the interactions between our microservices, ultimately ensuring the efficiency and reliability of our distributed system.

\begin{figure}
  \includegraphics[width=\linewidth]{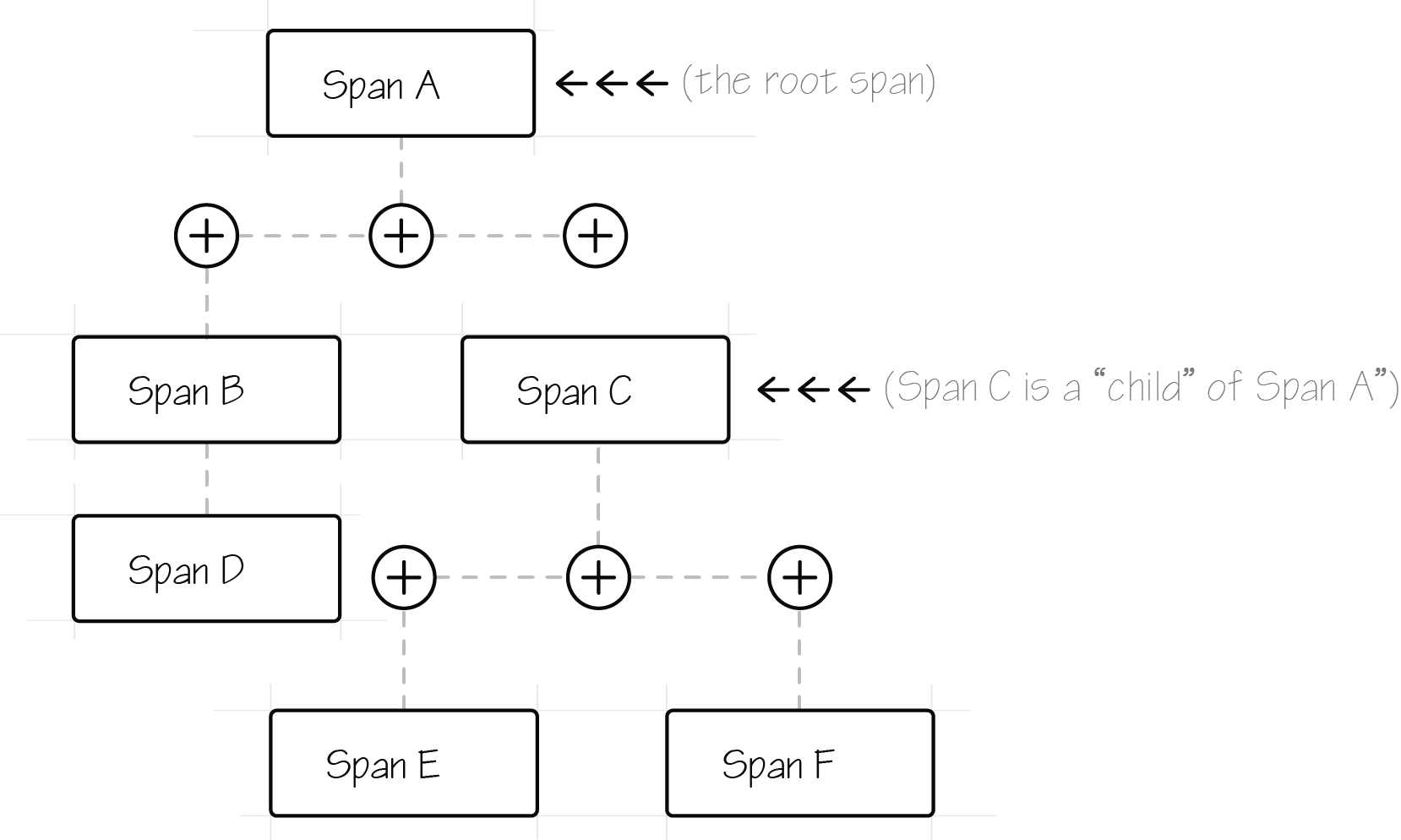}
  \caption{This example is taken from: \url{https://opentelemetry.io/docs/specs/otel/overview/}.}
  \label{fig:res2}
\end{figure}

Platformatic’s solution harnesses the power of data collected through OpenTracing spans to provide a robust mechanism for evaluating the risk associated with each pull request before it is merged into the codebase. OpenTracing, with its ability to capture detailed information about the execution of code and the interactions between microservices, forms the foundation of our risk assessment framework.
\par
As developers submit pull requests \footnote{\url{https://docs.github.com/en/pull-requests/collaborating-with-pull-requests/proposing-changes-to-your-work-with-pull-requests/about-pull-requests}}, our system leverages the traced data generated by OpenTracing to perform a comprehensive analysis. Each span, meticulously labeled with ``span-kind" to represent different types of operations, provides valuable insights into how the proposed code changes affect the system's behavior. By contextualizing these spans within the context of the pull request, our solution assesses potential breaking risks in the microservices network.
\par
This data-driven approach enables developers to catch the risk associated with individual changes before they land in the codebase. It empowers them with actionable insights, allowing them to proactively address any identified issues, optimize code, or make necessary adjustments to ensure the pull request has the minimum impact on the microservices network stability. Integrating OpenTracing data into our risk assessment process enables a more efficient and confident development workflow, ultimately leading to higher-quality software and reduced post-merge surprises.

\section{Conclusions}

In the ever-evolving realm of software development, adopting microservices architecture has brought a new era of agility and scalability. Yet, it has also presented organizations with an intricate challenge: managing the risks associated with constant change in this dynamic landscape. In this whitepaper, we have explored the critical need for a systematic and proactive approach to effectively measure and mitigate these risks.
\par
Our innovative system for measuring change risk within a microservices environment stands as a beacon of hope for organizations navigating this complexity. It offers a comprehensive solution, combining cutting-edge technology with data-driven insights, to empower decision-makers with the tools they need to steer their microservices ship through turbulent waters confidently.
\par
By delving into the intricacies of microservices, we have recognized that more than traditional risk assessment methods is required to address the unique challenges posed by this architecture. The granular nature of microservices, their interdependencies, and the sheer volume of change necessitate a specialized approach. Our system, with its advanced algorithms and real-time monitoring capabilities, identifies potential risks and quantifies them, enabling organizations to prioritize and act accordingly.
\par
Through real-world case studies and practical examples, we have witnessed the transformative impact of our system on organizations of various sizes and industries. It has enabled them to accelerate their innovation, reduce downtime, enhance system resilience, and, most importantly, build a culture of confidence in their microservices endeavors.
\par
As we look to the future, there are exciting avenues for further refinement and enhancement of our system. First and foremost, we recognize the importance of continuously improving the parameters used to calculate risk. This includes acknowledging the varying importance of different microservices within a system and the diverse values assigned to activities. By refining these parameters, we can tailor risk assessments to align with an organization's specific priorities and objectives.
\par
Additionally, we must remain vigilant in considering scenarios that still need to be considered. The technology landscape is ever-evolving, and new challenges and opportunities will undoubtedly emerge. Our commitment to innovation means staying at the forefront of these developments adapting our system to address novel risks and scenarios that may arise.
\par
In conclusion, the journey to mastering microservices risk management is ongoing, and our system represents a significant leap forward in that pursuit. We hope this whitepaper has provided valuable insights and inspired organizations to embrace this innovative approach. Together, we can navigate the intricate seas of microservices, charting a course to success, innovation, and resilience in an ever-changing digital world while continuously improving our methods and adapting to new challenges on the horizon.

\bibliographystyle{plain} 
\bibliography{refs} 

\end{document}